\documentstyle[osa,manuscript]{revtex}

\begin{document}
\titlepage
\title{%\vspace{-4cm} 
\hskip 9.0truecm {\normalsize FUB-HEP/96-1}\\
%\vspace{-0.4truecm}
%\hskip 11.20truecm  {\normalsize Submitted to Phys. Rev. D}\\
%\vspace {1.2truecm}
Production mechanisms and single-spin asymmetry 
for kaons in high energy hadron-hadron collisions} 
%\thanks{Supported in part by Deutsche Forschungsgemeinschaft (DFG:Me 470/7-1)}

\author {C. Boros, Liang Zuo-tang and Meng Ta-chung}
\address {Institut f\"ur Theoretische Physik,
Freie Universit\"at Berlin \\
Arnimallee 14, 14195 Berlin, Germany}
\maketitle

\begin{abstract}                % DON'T CHANGE THIS LINE

Direct consequences on kaon production 
of the picture proposed 
in a recent Letter and subsequent publications 
are discussed. 
Further evidence supporting 
the proposed picture is obtained.
Comparison with the data 
for the inclusive cross sections in unpolarized 
reactions is made. 
Quantitative results for the left-right asymmetry 
in single-spin processes are presented. 

\end {abstract}

%\narrowtext
%\twocolumn
\newpage

Single-spin asymmetry ($A_N)$ 
study has recently received much
attentions, both experimentally [1-7] and theoretically [8].
Striking experimental results have 
been obtained by FNAL E704
Collaboration [3-7] at high energy (200 GeV/c) 
for mesons as well as for Lambda hyperon. 
Large left-right asymmetries 
have been observed 
in the fragmentation region 
of the polarized beam particles. 
The data shows that the asymmetry 
is not only 
different for reactions using different 
polarized beams 
but also  
different for the production of different kinds of hadrons. 

In a recent Letter [10] and 
the subsequent publications [9-13], 
we suggested that these 
observed asymmetries 
are due to the orbital motion of the valence
quarks of the polarized projectiles. 
Theoretical arguments and experimental indications
supporting the proposed picture have been given; 
and the characteristic features
of the proposed picture have been discussed. 
It has been pointed out in particular
that measurements of such single-spin asymmetries 
are not only extremely useful in studying the 
spin structure of the nucleon but also 
very helpful in studying the production mechanisms 
of hadrons in high energy hadron-hadron collisions. 
It has been shown, in this connection, that 
only part of the observed hadrons are 
direct formation (fusion) products  
of orbiting valence quarks of the projectile
and anti-sea-quarks associated with the target. 
The observed left-right asymmetry 
in particular its $x_F$-dependence reflects 
the interplay between 
the contribution of the direct formation mechanism 
and the non-direct-formation part.
(Here, $x_F\equiv 2p_\parallel /\sqrt{s}$,  
$p_\parallel $ is the longitudinal 
momentum of the produced hadron, 
and $s$ is the center of mass energy squared 
for the colliding hadron system.)
Single-spin asymmetry measurements 
can in particular differentiate 
these two kinds of contributions 
since only the former 
depends on the polarization of the projectile.
Direct consequences of such a picture 
for the production of pions and $\Lambda$ have been 
studied in detail in single-spin, as well as in 
unpolarized, hadron-hadron collision processes. 
Good agreements between experiments [2-7] 
and theory [9-13] have been found.
It is then natural to ask: 
What do we expect to see for the production of other 
kinds of mesons such as kaons? 
How are they compared with experiments?
A qualitative discussion on the left-right 
asymmetry for $K$ production 
has already been made in [12].  
The purpose of this brief report is to 
present the detailed study of the 
direct consequences of this picture on the production of 
kaons and anti-kaons in unpolarized reactions 
and to present the quantitative calculations 
for the left-right asymmetry in single-spin processes. 
This is of particular interest for the following reasons: 
First, preliminary data for the left-right asymmetry for $K_s^0$  
is now available from FNAL E704 Collaboration [7].  
They are consistent with our qualitative predictions 
in [12]. 
Second, for kaons, direct fusion of the valence quarks of 
the proton-projectile with suitable anti-sea-quarks of the targets 
does not contribute to the production of 
$K^-$ and $\bar K^0$.  
This means, for such kaons, we have only the 
contributions from the non-direct-formation parts.
Hence, the non-direct-formation part 
for kaon production can be 
determined unambiguously even in experiments 
using unpolarized projectiles and unpolarized targets. 
Comparison of the results obtained from 
the proposed picture with the data 
from both unpolarized and polarized experiments  
provides further quantitative 
tests of the picture.

We begin our discussions by recollecting some of the 
formulae which are particularly useful here. 
We recall that the left-right asymmetry $A_N(x_F,h|s)$ 
for hadron production in single-spin process 
$p(\uparrow )+p(0)\to h+X$ is defined [1-7] as 
the ratio of the difference and the sum 
of $N(x_F,h|s,\uparrow )$ and 
$N(x_F,h|s,\downarrow )$. 
Here, $h$ stands for a hadron which can be a pion, a kaon 
or a hyperon; and 
$N(x_F,h|s,\uparrow )$
is the number-density of $h$ observed in a given kinematic 
region $R$ (for example, with transverse momentum 
$p_\perp \ge 0.7$GeV/c and
in a given acceptance solid angle 
on the left-hand-side looking down stream
in the above-mentioned experiments [2-7]) 
in $p(\uparrow )+p(0)\to h +X$ at total c.m.s.-energy $\sqrt {s}$ 
using upwards transversely polarized proton projectile $p(\uparrow )$ 
and unpolarized proton target $p(0)$.
$N (x_F, h|s,\downarrow)$ 
is the corresponding density function 
for such hadrons observed in $p(\downarrow )+p(0)\to h+X$. 
We consider now the cases where $h=M$ is a meson $M$ 
and denote by $D(x_F,M,+|s,\uparrow )$ the 
number-density for those mesons $M$ 
which are directly formed by the
valence quarks polarized 
in the same direction as the 
transversely polarized projectile proton, 
and $D(x_F,M,-|s,\uparrow )$ the 
corresponding number density for mesons 
formed by the valence quarks polarized in 
the opposite direction as the projectile proton.
We recall that these $D$'s can be expressed [10] as 
the following integrals,
\begin{equation}
D(x_F,M,\pm \ | \ s,tr) = \sum _{q_v,\bar q_s}
\int  dx^P dx^T q_v^{\pm}(x^P | s,tr)
\bar q_s(x^T | s)K(x^P,q_v;x^T, \bar q_s| x_F, M,s). 
\end{equation}
Here $q^\pm _v(x|s,tr)$
is the distribution of the valence quarks
polarized in the same or in the opposite direction of the
transversely polarized proton, and $\bar q_s(x|s)$
is the spin-averaged sea-quark distribution;
$K(x^P,q_v;x^T, \bar q_s| x_F, M,s)$ 
is the probability density for a valence quark 
of flavor $q_v$ with fractional momentum fraction $x^P$ 
to combine directly with an anti-sea-quark of 
flavor $\bar q_s$ with fractional momentum $x^T$ 
to form a meson $M$ of fractional momentum $x_F$.
We recall also 
whether, if yes how much, 
the $K$-function depends on
the dynamical details of this direct formation process
is something we do not know a priori.
But, what we do know is that 
this function has to guarantee 
the validity of all
the relevant conservation laws.
Hence, it contains, in practice, 
as factors a product of Kronecker delta's and
Dirac $\delta $-functions, where every
one is associated with a given quantum number.
This implies, in particular, that the simplest choice of the
corresponding $K$-function 
for $K^+=(u\bar s)$ or $K^0=(d\bar s)$ production 
is the following: 
\begin{equation}
K(x^P,q_v;x^T, \bar q_s| x_F, K^+,s)=\kappa _K 
\delta _{q_v,u} \delta _{\bar q_s,\bar s} 
\delta (x^P-x_F) \delta (x^T-x_0/x_F), 
\end{equation}
\begin{equation}
K(x^P,q_v;x^T, \bar q_s| x_F, K^0,s)=\kappa _K 
\delta _{q_v,d} \delta _{\bar q_s,\bar s} 
\delta (x^P-x_F) \delta (x^T-x_0/x_F). 
\end{equation}
where $\kappa_K$ is a constant;  
the two Dirac-$\delta$-functions 
come from the energy and momentum
conservation which requires
$x^P\approx x_F$ and $x^T\approx x_0/x_F$ where $x_0=m^2/s$
($m$ is the mass of the produced meson).  
We thus obtain, 
\begin{equation}
D(x_F,K^+,\pm|s,tr)=\kappa _K u^\pm_v(x_F|s,tr)\bar s_s(x_0/x_F|s),
\end{equation}
\begin{equation}
D(x_F,K^0,\pm|s,tr)=\kappa _K d^\pm_v(x_F|s,tr)\bar s_s(x_0/x_F|s).
\end{equation}

According to the proposed picture [9,10], 
the difference 
$\Delta N(x_F,M|s,tr)
\equiv N(x_F,M|s,\uparrow\nobreak)- N(x_F,M|s,\downarrow )$ 
comes only from those mesons that are directly 
formed through the fusion of 
the orbiting valence quarks of the polarized projectile 
with suitable anti-sea-quark associated with the target, i.e., 
$\Delta N(x_F,M|s,tr)=C\Delta D(x_F,M|s,tr)$.  
[Here $0<C<1$ is a constant, and
$\Delta D(x_F,M|s,tr)\equiv D(x_F,M,+|s,tr)-D(x_F,M ,-|s,tr)$.]
The sum of them is nothing else but two times the
corresponding number density $N(x_F,M|s)$ in reactions 
using unpolarized projectiles and unpolarized targets,  
\begin{equation}
N(x_F,M|s)=N_0(x_F,M|s) +D(x_F,M|s), 
\end{equation}
where $D(x_F,M|s)\equiv [D(x_F,M,+|tr,s)+D(x_F,M,-|tr,s)]/2$.[14] 
We have therefore, 
\begin{equation}
A_N(x_F,M|s)={C\Delta D(x_F,M|s,tr) \over 2[N_0(x_F,M|s)+D(x_F,M|s)]},
\end{equation}
It can easily be seen that 
$N_0(x_F,M|s)$ ----- especially the interplay 
between this quantity and the corresponding $D(x_F,M|s)$
----- plays a key role 
in understanding the $x_F$-dependence of $A_N(x_F,M|s)$.

The direct consequences of these equations 
for pion production have been discussed in detail 
in [9-11] and they are in good agreement with experiments.
Now let us consider $K$-production and see 
what they tell us.
First, in unpolarized collision processes 
$p(0)+p(0)\to K+X$, 
since there is no contribution from the direct 
fusion of the valence quarks of the projectile with 
suitable anti-sea-quarks of the target 
to $K^-=(\bar us)$ and $\bar K^0=(\bar ds)$, 
we obtain from Eq.(6) that,
\begin{equation}
N(x_F,K^+|s)=N_0(x_F,K^+|s) +D(x_F,K^+|s), 
\end{equation}
\begin{equation}
N(x_F,K^0|s)=N_0(x_F,K^0|s)+D(x_F,K^0|s), 
\end{equation}
\begin{equation}
N(x_F,K^-|s)=N_0(x_F,K^-|s), \phantom{+D(x_F,K^0|s)} 
\end{equation}
\begin{equation}
N(x_F,\bar K^0|s)=N_0(x_F,\bar K^0|s),\phantom{+D(x_F,K^+|s)}  
\end{equation}
where $D(x_F,K^+|s)=\kappa_K u_v(x_F|s)\bar s_s(x_0/x_F|s)/2$,  
and $D(x_F,K^0|s)=\kappa_K d_v(x_F|s)\bar s_s(x_0/x_F|s)/2$.  
$N_0$ comes from the interactions of the sea
(the sea quarks, the sea antiquarks and the gluons)  
of the projectile with that of the target
and is therefore expected to be 
isospin invariant and to be the 
same for particle and anti-particle.  
This implies, in particular for $K$-production, 
that, 
$N_0(x_F,K^+|s)=N_0(x_F,K^0|s)
=N_0(x_F,K^-|s)=N_0(x_F,\bar K^0|s)
\equiv N_0(x_F,K|s)$. 
Hence, we obtain the following relations 
between the number densities 
(or the corresponding differential cross sections) 
for $K$ produced in $p(0)+p(0)\to K+X$:
\begin{equation}
N(x_F,K^-|s)=N(x_F,\bar K^0|s)=N_0(x_F,K|s),
\end{equation}
\begin{equation}
N(x_F,K^+|s)=N(x_F,K^-|s)+\kappa _K u_v(x_F|s)\bar s_s(x_0/x_F|s)/2,
\end{equation}
\begin{equation}
N(x_F,K^0|s)=N(x_F,K^-|s)+\kappa _K d_v(x_F|s)\bar s_s(x_0/x_F|s)/2,
\end{equation}
Recall that $K_S^0\approx (K^0+\bar K^0)/\sqrt{2}$ and 
$K_L^0\approx (K^0-\bar K^0)/\sqrt{2}$, we have 
\begin{equation}
N(x_F,K^0_S|s)=
N(x_F,K^0_L|s)=N(x_F,K^-|s)+\kappa _K d_v(x_F|s)\bar s_s(x_0/x_F|s)/4.
\end{equation}
These are direct consequences of the proposed picture 
which can be tested by unpolarized experiments. 
For example, Eq.(13) implies that 
$N(x_F,K^+|s)$ is determined completely by 
$N(x_F,K^-|s)$ and the spin averaged quark distribution functions. 
The only parameter is the constant 
$\kappa _K$ which can be determined 
by fitting one data point.  
In Fig.~1, we compare such results 
with data [15,16] 
and we see that the agreement is indeed very good.

Next, we consider the single-spin process 
$p(\uparrow)+p(0)\to K+X$.
It follows from Eqs.(4),(5),(7) and (13-14) that, 
\begin{equation}
A_N(x_F,K^+|s)={C\kappa_K\Delta u_v(x_F|s,tr) \bar s_s(x_0/x_F|s) 
\over 2N(x_F,K^-|s)+\kappa_K u_v(x_F|s) \bar s_s(x_0/x_F|s) },
\end{equation}
\begin{equation}
A_N(x_F,K^0|s)={C\kappa_K\Delta d_v(x_F|s,tr) \bar s_s(x_0/x_F|s) 
\over 2N(x_F,K^-|s)+\kappa_K d_v(x_F|s) \bar s_s(x_0/x_F|s) },
\end{equation}
and $A_N(x_F,K^-|s)=A_N(x_F,\bar K^0|s)=0$. 
These are the same results as those in [12] 
[see Eqs. (22) and (23) there]
with $N_0(x_F,K|s)$ replaced by $N(x_F,K^-|s)$. 
From them, we expected 
that $A_N(x_F,K^+|s)$ is similar to $A_N(x_F,\pi^+|s)$, and 
that $A_N(x_F,K^0|s)$ is similar to $A_N(x_F,\pi^-|s)$. 
Since both the $K_S^0$ and $K_L^0$ are linear combinations 
of $K^0$ and $\bar K^0$, the left-right asymmetry is the same 
for them, and it is given by
\begin{equation}
A_N(x_F,K^0_S|s)={C\kappa_K\Delta d_v(x_F|s,tr) \bar s_s(x_0/x_F|s) 
\over 4N(x_F,K^-|s)+\kappa_K d_v(x_F|s) \bar s_s(x_0/x_F|s)}.
\end{equation}
Hence $A_N(x_F,K^0_S|s)$ should have the same sign 
as $A_N(x_F,\pi^-|s)$. This is confirmed by the 
preliminary data of E704 Collaboration [7].
Now we can use the parameterization of $N(x_F,K^-|s)$ 
and those for the quark distribution functions 
to calculate these $A_N$ quantitatively.
The results are given in Fig.2. 

In this connection, 
it may be particularly interesting to note the following. 
If we use, instead of transversely polarized proton beam,  
transversely polarized anti-proton beam, 
we have, 
\begin{equation}
A_N^{\bar p(\uparrow)p}(x_F,K^-|s)=
{C\kappa_K\Delta \bar u_v^{\bar p}(x_F|s,tr) s_s(x_0/x_F|s) 
\over 2N(x_F,K^-|s)+\kappa_K \bar u_v^{\bar p}(x_F|s) s_s(x_0/x_F|s) },
\end{equation}
\begin{equation}
A_N^{\bar p(\uparrow)p}(x_F,K^0_S|s)=
{C\kappa_K\Delta \bar d_v^{\bar p}(x_F|s,tr) s_s(x_0/x_F|s)
\over 4N(x_F,K^-|s)+\kappa_K \bar d_v^{\bar p}(x_F|s) s_s(x_0/x_F|s) },
\end{equation}
and $A_N^{\bar p(\uparrow)p}(x_F,K^+|s)=0$. 
Using $\Delta \bar q_v^{\bar p}(x|s,tr)=\Delta q_v(x|s,tr)$
and $\bar q_v^{\bar p}(x|s)=q_v(x|s)$, 
we obtain that 
\begin{equation}
A_N^{\bar p(\uparrow)p}(x_F,K^-|s)=A_N(x_F,K^+|s),
\end{equation}
\begin{equation}
A_N^{\bar p(\uparrow)p}(x_F,K^0_S|s)=A_N(x_F,K^0_S|s).
\end{equation}
Eq.(21) is the same as what we obtained in [12]. 
Eq.(22) is a direct consequence 
of what we had in [12] for $K^0$ and $\bar K^0$.  
The latter is also confirmed 
by the preliminary data of E704 Collaboration [7].

In summary, 
direct consequence on $K$ production 
of the picture proposed 
in a recent Letter [10] 
and subsequent publications [9-13] 
are discussed. 
Further evidences supporting the proposed picture 
are obtained in unpolarized, 
as well as in single-spin, processes. 
Quantitative predictions have been made 
for the left-right 
asymmetry for kaon production 
in single spin proton-proton or anti-proton-proton 
collision processes. 
The results are in agreement with the preliminary  
data of FNAL E704 Collaboration [7] and can be 
further tested by experiments in the future. \\

This work was supported in part by Deutsche
Forschungsgemeinschaft (DFG: Me \mbox{470/7-1)}.

\newpage

\noindent
Figure captions

\vspace {0.5truecm}

\noindent
Fig. 1. 
Inclusive invariant cross section 
$Ed^3\sigma/d^3p$ for 
$p(0)+p(0)\to K^++X$ as a function of $x_F$ at 
$p_\perp =0.75$ GeV/c and the ISR energies 
is shown as the sum of the following two parts: 
(1) the isospin-independent non-direct-formation part
which is taken as the same as 
$Ed^3\sigma/d^3p$ for $p(0)+p(0)\to K^-+X$ 
[parameterized as $N(1-x_F)^3exp(-10x_F^3)$, 
shown by the dashed curve], 
(2) the corresponding flavor-dependent 
direct formation part
$\kappa _K u_v(x_F)\bar s_s(x_0/x_F) x_F/2$ 
(shown by the dashed-dotted curve).
Data are taken from [15] and [16]. 
Those at larger $x_F$ are from [15] and they are for 
$p_\perp =0.75$ GeV/c. 
Those at lower $x_F$ are from [16] and  
they are for $p_\perp =0.8$ GeV/c.

\noindent
Fig. 2.
Left-right asymmetry $A_N$ for  
$p(\uparrow )+p(0)\to K^++X$ 
and that for 
$p(\uparrow )+p(0)\to K^0_S+X$ 
as functions of $x_F$ at $200$ GeV/c.  
The data points are the preliminary results 
from FNAL E704 Collaboration[7].


\begin{references}
\bibitem{14.} Please note that there is a misprint 
             in the definition of $D(x_F,M|s)$ and the corresponding 
             expressions for pions and kaons in [12]. 
             The factor ``/2'' is missing there.
\end{references}
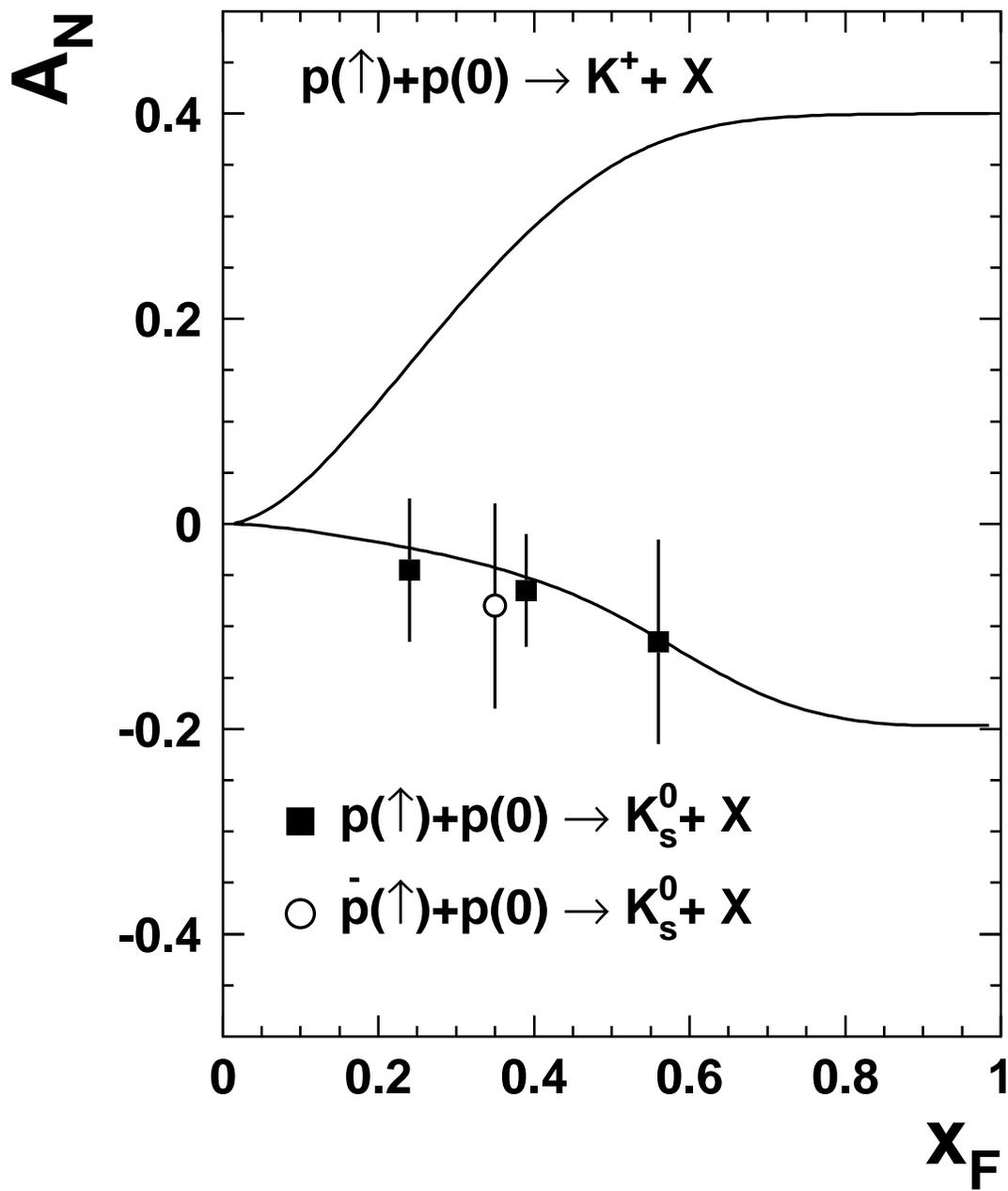
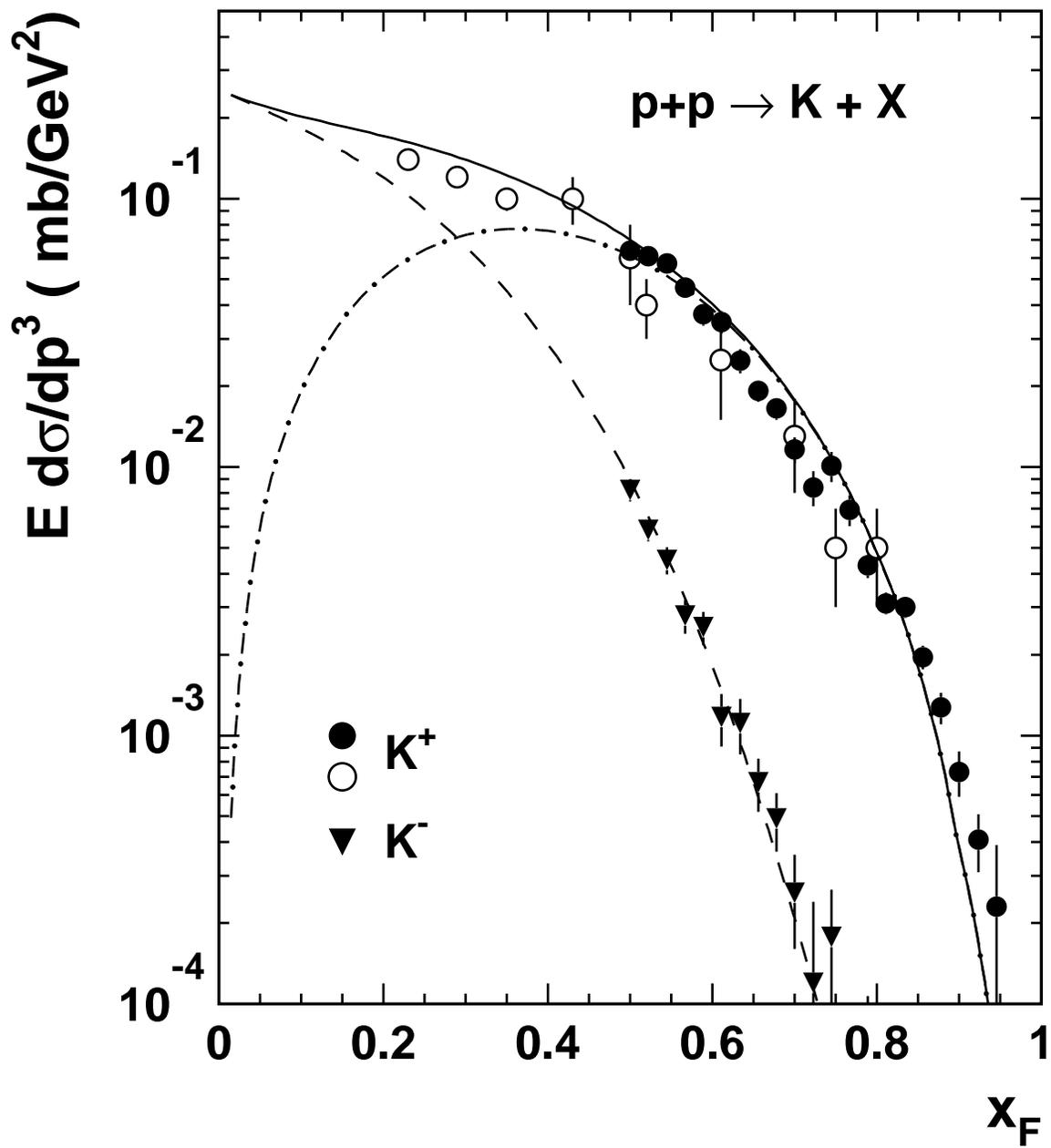
\end{document}